\documentclass[]{aa}
\pdfoutput=1                            
\usepackage[varg]{txfonts}

\usepackage{amsmath}
\usepackage{amsfonts}
\usepackage{amssymb}
\usepackage{wasysym}
\usepackage{graphicx}

\usepackage{natbib}
\usepackage{epsfig}
\usepackage{xspace}
\usepackage{booktabs}
\usepackage{dcolumn}
\usepackage[para]{footmisc}
\usepackage{placeins}
\usepackage{multicol}

\usepackage[switch]{lineno}

\usepackage[utf8]{inputenc}
\usepackage[english]{babel}

\newcommand{\emth}[1]{\ensuremath{#1}\xspace}
\newcommand{\parm}[1]{\ensuremath{\theta_\mathrm{#1}}\xspace}

\newcommand{\covmat}{\emth{\boldsymbol{\Sigma}}}

\newcommand{\npb}{\emth{n_\mathrm{pb}}}
\newcommand{\nph}{\emth{n_\mathrm{ph}}}
\newcommand{\nip}{\emth{n_\mathrm{\mu}}}

\newcommand{\gcm}{\emth{\mathrm{g\,cm^{-3}}}}

\newcommand{\logg}{log {\it g}}

\newcommand{\ud}{\ensuremath{\mathrm{d}}}

\newcommand{\llh}{\ensuremath{\ln P}}
\newcommand{\iid}{i.i.d.\xspace}

\newcommand{\pvec}{\ensuremath{\boldsymbol{\theta}}\xspace}
\newcommand{\Dlc}{\ensuremath{\vec{D_{\mathrm{LC}}}}\xspace}

\newcommand{\Dld}{\ensuremath{\vec{D_{\mathrm{LD}}}}\xspace}

\newcommand{\NP}[1]{\ensuremath{N(#1)}\xspace}

\newcommand{\prr}{\parm{k}}           		
\newcommand{\pper}{\parm{p}}          		
\newcommand{\pa}{\parm{a}}            		
\newcommand{\pec}{\parm{e}}           		
\newcommand{\pom}{\parm{\omega}}      		
\newcommand{\pin}{\parm{i}}           		
\newcommand{\ptc}{\parm{tc}}
\newcommand{\ldu}{\parm{u}}
\newcommand{\ldv}{\parm{v}}
\newcommand{\prho}{\parm{\rho}} 

\newcommand{\pbl}{\parm{B}}			
\newcommand{\pext}{\parm{x}}			
\newcommand{\pip}{\parm{b}}	                

\newcommand{\kobs}{\emth{k_\mathrm{obs}}}
\newcommand{\teff}{\emth{T_\star}}
\newcommand{\kreal}{\emth{k_\mathrm{real}}}
\newcommand{\tcont}{\emth{T_\mathrm{c}}}
\newcommand{\tspot}{\emth{T_\mathrm{s}}}

\newcommand{\rstar}{\emth{R_\star}}

\newcommand{\changed}[1]{\textbf{#1}}
\renewcommand{\changed}[1]{{#1}}

\begin{document}
\title{The GTC exoplanet transit spectroscopy survey II}
\subtitle{An overly-large Rayleigh-like feature for exoplanet TrES-3b}

\author{
 Parviainen, H.\inst{\ref{ioxford}}
 \and Pall\'e, E.\inst{\ref{iiac},\ref{iull}} 
 \and Nortmann, L.\inst{\ref{igott}}
 \and Nowak, G.\inst{\ref{iiac},\ref{iull}} 
 \and Iro, N.\inst{\ref{iham}} 
 \and Murgas, F.\inst{\ref{igr},\ref{icnrs}}
 \and Aigrain, S.\inst{\ref{ioxford}}
}

\institute{
 Sub-department of Astrophysics, Department of Physics, University of Oxford, Oxford, OX1 3RH, UK\label{ioxford}
\and Instituto de Astrof\'isica de Canarias (IAC), E-38200 La Laguna, Tenerife, Spain\label{iiac}
\and Dept. Astrof\'isica, Universidad de La Laguna (ULL), E-38206 La Laguna, Tenerife, Spain\label{iull}
\and Institut f\"ur Astrophysik, Georg-August-Universit\"at, Friederich-Hund-Platz 1, 37077 G\"ottingen, 
Germany\label{igott}
\and Theoretical Meteorology group, Klimacampus, University of Hamburg, Grindelberg 5, 20144 Hamburg, Germany
\label{iham}
\and Univ. Grenoble Alpes, IPAG, F-38000 Grenoble, France\label{igr}
\and CNRS, IPAG, F-38000 Grenoble, France\label{icnrs}
}
\date{Received ; accepted}

\abstract{}
  {\changed{We set to search for Rayleigh scattering and K and Na absorption signatures from the atmosphere} of TrES-3b 
using ground-based transmission spectroscopy covering the wavelength range from 530 to   950~nm as observed with 
OSIRIS@GTC.}
  {Our analysis is based on a Bayesian approach where the light curves covering a set of given passbands are fitted
  jointly with PHOENIX-calculated stellar limb darkening profiles. \changed{The analysis is carried out assuming both
  white and red -- temporally correlated -- noise, with two approaches (Gaussian processes and
  divide-by-white) to account for the red noise.}}
  {\changed{An initial analysis reveals a transmission spectrum that shows a strong Rayleigh-like increase in extinction
towards
  the blue end of the spectrum, and enhanced extinction around the K~I resonance doublet near 767~nm. However, 
  the signal amplitudes are significantly larger than expected from theoretical considerations. A detailed analysis
  reveals that the K~I-like feature is entirely due to variability in the telluric O$_2$ absorption, but the
  Rayleigh-like feature remains unexplained.}
} {}


\titlerunning{}
\authorrunning{}

\maketitle

\section{Introduction}
\label{sec:introduction}

Transmission spectroscopy offers a powerful means for the characterisation of transiting exoplanet atmospheres.
Measuring how the transit depth changes as a function of wavelength allows us to probe the existence and abundance of
different atmospheric species -- each with their wavelength-dependent extinction features -- in the planet's atmosphere
\citep{Seager2000,Brown2001a}. However, the variations in the transit depth are small, and high-altitude clouds can mask
them altogether, leading to a flat transmission spectrum \citep{Kreidberg2013,Berta2011}. Further, atmospheric
extinction is not the only source of  wavelength-dependent features in transmission spectra, but stellar sources, such
as star spots \citep{Ballerini2012}, plages \citep{Oshagh2014}, and flux contamination from an unresolved source can
imprint features that can be difficult to disentangle from the atmospheric effects.

Since the colour variations in the transit depth are small -- even in the absence of clouds -- high-precision
spectroscopic
time series are required for meaningful analyses. Transmission spectroscopy has been most successful from space
\citep[][etc.]{Charbonneau2002,Sing2011,Gibson2012}, but the recent developments in observing techniques and modern data
analysis methods have led to improvements in the precision that can be achieved from the ground. Simultaneous 
measurements of the target star and several comparison stars, similar to relative
photometry~\citep{Bean2010a,Gibson2012a}, and the use of Gaussian processes, have facilitated the correction of
systematics by allowing for the robust modelling of correlated noise -- including time correlation and correlations with
auxiliary measurements such as seeing -- in model-independent fashion~\citep{Roberts2013,Gibson2011a,Rasmussen2006,
Murgas2014}.

We have observed a spectroscopic time series of a transit of \object{TrES-3b}, a massive hot Jupiter around a metal poor
V=12.4 G-star on a 1.3~d orbit (see Table~\ref{tbl:star} and \citealt{O'Donovan2007}). The observations were carried out
 with the OSIRIS spectrograph \citep[Optical System for Imaging and low-Intermediate-Resolution Integrated
Spectroscopy;][]{Sanchez2012} installed in the 10.4~m Gran Telescopio CANARIAS (GTC) in La Palma island. The
observations cover the spectral range from 500~to~900~nm, probing the planet's atmosphere for a possible Rayleigh
scattering signal in the blue end of the spectrum, and the visible-light extinction features of the K~I and Na~I
resonance doublets at 767~nm and 589.4~nm, respectively.

\changed{We detail our observations and data reduction procedures in (§\ref{sec:observations}), the theoretical basis 
and the numerical methods in (§\ref{sec:theory}), broadband (white) light curve modelling in 
(§\ref{sec:broadband_analysis}), transmission spectroscopy in (§\ref{sec:transmission_spectroscopy}), and finally 
conclude with a discussion of the results (§\ref{sec:conclusions}). The analysis and the raw data are publicly 
available on Github,$\!$\footnote{\url{github.com/hpparvi/Parviainen-2015-TrES-3b-OSIRIS}} as an easy-to-follow set of 
IPython notebooks and Python codes, to help with reproducibility of the study.}

\begin{table}[t]    
  \caption{Identifiers for TrES-3 with its coordinates and magnitudes.}
  \centering
  \begin{tabular*}{\columnwidth}{@{\extracolsep{\fill}} lll}
  \toprule\toprule
  \multicolumn{1}{l}{\emph{Main identifiers}}     \\
  \midrule              
  GSC~ID          & 03089-00929       \\
  USNO-A2~ID      & 45017453          \\
  2MASS~ID        & J17520702+3732461 \\
  WASP~ID         & 1SWASP J175207.01+373246.3 \\
  \midrule               
  \multicolumn{2}{l}{\emph{Equatorial coordinates}}     \\
  \midrule            
  RA \,(J2000)      & $17^h\,52^m\,07\fs02$              \\
  Dec (J2000)       & $+37\degr\,32\arcmin\,46\farcs2$  \\
  \midrule              
  \multicolumn{3}{l}{\emph{Magnitudes}} \\
  \midrule              
  \centering
  Filter & Magnitude       & Error  \\
  \midrule                
  $B$  & 13.114 & 0.009 \\
  $V$  & 12.402 & 0.006 \\
  $R$  & 12.060 & - \\
  $I$  & 11.603 & 0.010 \\
  $J$  & 11.015 & 0.022 \\
  $H$  & 10.655 & 0.030 \\
  $K$  & 10.608 & 0.028 \\
  \bottomrule
  \end{tabular*}
  \label{tbl:star}  
\end{table}

\section{Observations} 
\label{sec:observations}
\subsection{Overview}
\label{sec:observations:overview}

Observations were taken with OSIRIS@GTC on the night of 8~July~2014, during a transit of TrES-3b. A spectroscopic time 
series was taken in staring mode, from 1:27 to 4:00 UT, a total of 2.55 hours, starting 30~min before the ingress, and 
finishing 40~min after the egress, with a total of 255$\times$12~s exposures. Observing conditions were good, with 
median seeing at 0.86\arcsec, and the airmass varying from 1.06 to 1.56.

\begin{figure}
 \centering
 \includegraphics[width=\columnwidth]{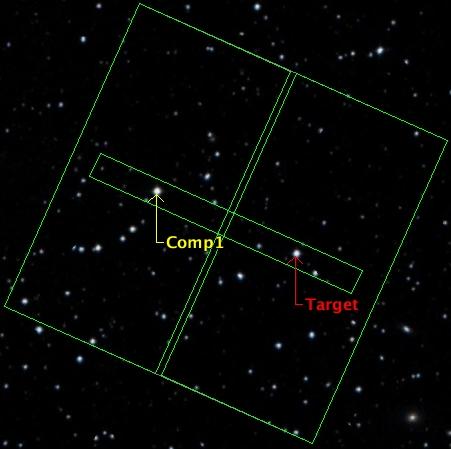}
 \caption{Finding chart showing the field of view of OSIRIS and the slit position within the field (green boxes).  The 
  target star (TrES-3, right) and the comparison star (left) are also marked.  Both OSIRIS CCDs were used for the  
  observations, with one star located in each of the CCDs.} 
 \label{fig:finding_chart}
\end{figure}

OSIRIS contains two 2048$\times$4096 pixel E2V CCDs, which were used in the 2$\times$2 binning mode. The observations
were carried out using grism R1000R with a 40\arcsec-wide slit, with the target and a comparison star both located in
the slit. We chose \object{TYC~3089-995-1} as the comparison star, a star with a similar colour to TrES-3 and located at
a distance of 3.93\arcmin{} from it. The position angle of the reference star with respect to the target was~66\degr.
The two stars were positioned equidistantly from the optical axis, close to the centre of each CCD, as shown in
Fig.~\ref{fig:finding_chart}. The slit also includes several fainter stars, but these were found to be too faint to be
useful in the data reduction. One of the faint stars (\object{2MASS~17520839+3732378}, see
Sect.~\ref{sec:data:contamination}) is within a very short projected distance from TrES-3, and is included inside the
aperture used to calculate the light curves. While the contaminating star is faint, its effect needs to be accounted for
in the analysis.

\subsection{Generation of the spectra}
\label{sec:observations:data_reduction}

\begin{figure}
 \centering
 \includegraphics[width=\columnwidth]{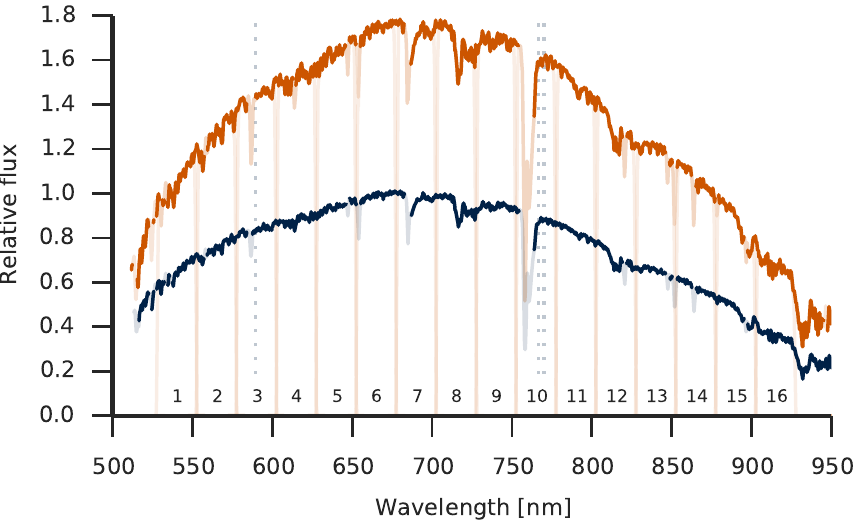}
 \caption{Sky-subtracted, wavelength-calibrated, spectra for TrES-3 (dark blue line, below) and the simultaneously
observed comparison star (orange line, above), both normalised with an arbitrary but common normalisation factor. The
parts masked out during the generation of spectrum-masked light curves (see below) are shown with a lighter shade, and
the light-orange vertical lines show the ranges used to generate the 16 narrow-band light curves. The dotted vertical
lines show the Ka~I and Na~I resonance doublets.}
 \label{fig:spectra}
\end{figure}

The 2D images were reduced to 1D spectra following the normal procedures for long-slit spectroscopy using routines
within the IRAF\footnote{IRAF is distributed by the National Optical Astronomy Observatory, which is operated by the
Association of Universities for Research in Astronomy (AURA) under cooperative agreement with the National Science
Foundation.} environment. The raw spectra were bias corrected by subtracting the median-combined bias exposures and flat
fielded by dividing by the normalised continuum lamp. Several apertures were tested for the optimal extraction, and the
aperture that produced the white light curve with the smallest root mean square (RMS) scatter was finally chosen. For
TrES-3b the aperture width was 50 binned pixels, which corresponds to 12.7\arcsec{} on the detector (8.2 to
11.8 times the raw seeing during the observations). For TYC~3089-995-1 the aperture width was chosen to be 40 binned
pixels, i.e. 10.16\arcsec{} on the detector (6.5 to 9.6 times the raw seeing). Wavelength calibration was performed
using the HgAr, Xe, Ne lamps for 1\arcsec-wide slit and a Chebyshev function fit of order 6, providing RMS better than
0.04~{\AA}. Final spectra were not corrected for instrumental response nor flux calibrated. Fig.~\ref{fig:spectra} shows
the example spectra for the two stars used in time series analysis.

\subsection{Light curve generation}
\label{sec:observations:ligh_curves}
\subsubsection{Basic light curve set without spectrum masking}

\begin{figure}
 \centering
 \includegraphics[width=\columnwidth]{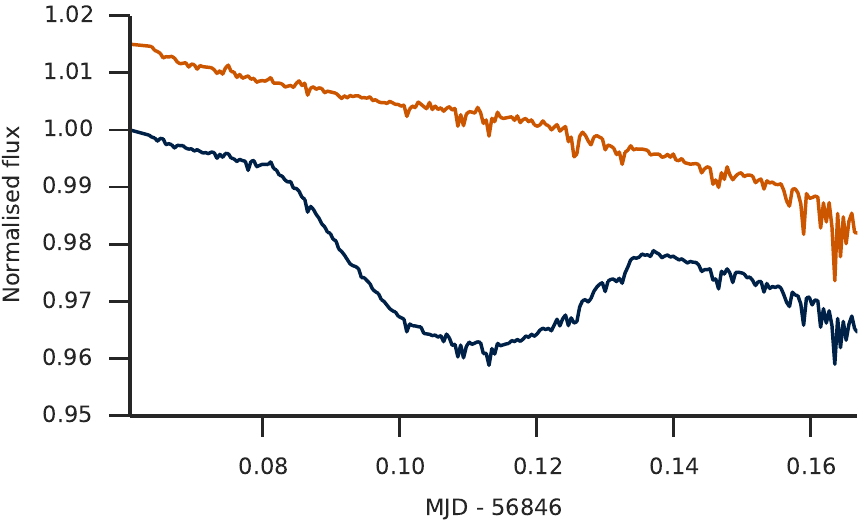}
 \caption{Raw broadband light curves for TrES-3 (below) and the simultaneously observed comparison star 
(above). The light curves are normalised to their first datapoint, and the comparison star's light curve is shifted 
vertically for clarity.} 
\label{fig:raw_white_fluxes}
\end{figure}

\changed{Two sets of light curves, with and without spectrum masking, were created from the raw spectra for each star. 
First, we carried out an initial analysis using a basic (unmasked) light curve set with a broadband light curve 
(Fig.~\ref{fig:raw_white_fluxes}) integrating the flux over the whole usable spectral range from 530~to~930~nm, and 16 
narrow-band ($\sim$25~nm) light curves integrated over the spectral ranges shown in Fig.~\ref{fig:spectra}.}

\subsubsection{Light curve set with spectrum masking}
\changed{We realised during the initial analysis that the spectral regions corresponding to the cores of telluric
absorption bands added a significant amount of noise to the narrow-band light curves. This was especially the case with 
the deep telluric O$_2$ absorption band near 760~nm. This motivated us to create a second set of light curves to 
assess how these spectrum regions affected the light curves, and, finally, our parameter estimates.}

\begin{figure*}
 \centering
 \includegraphics[width=\textwidth]{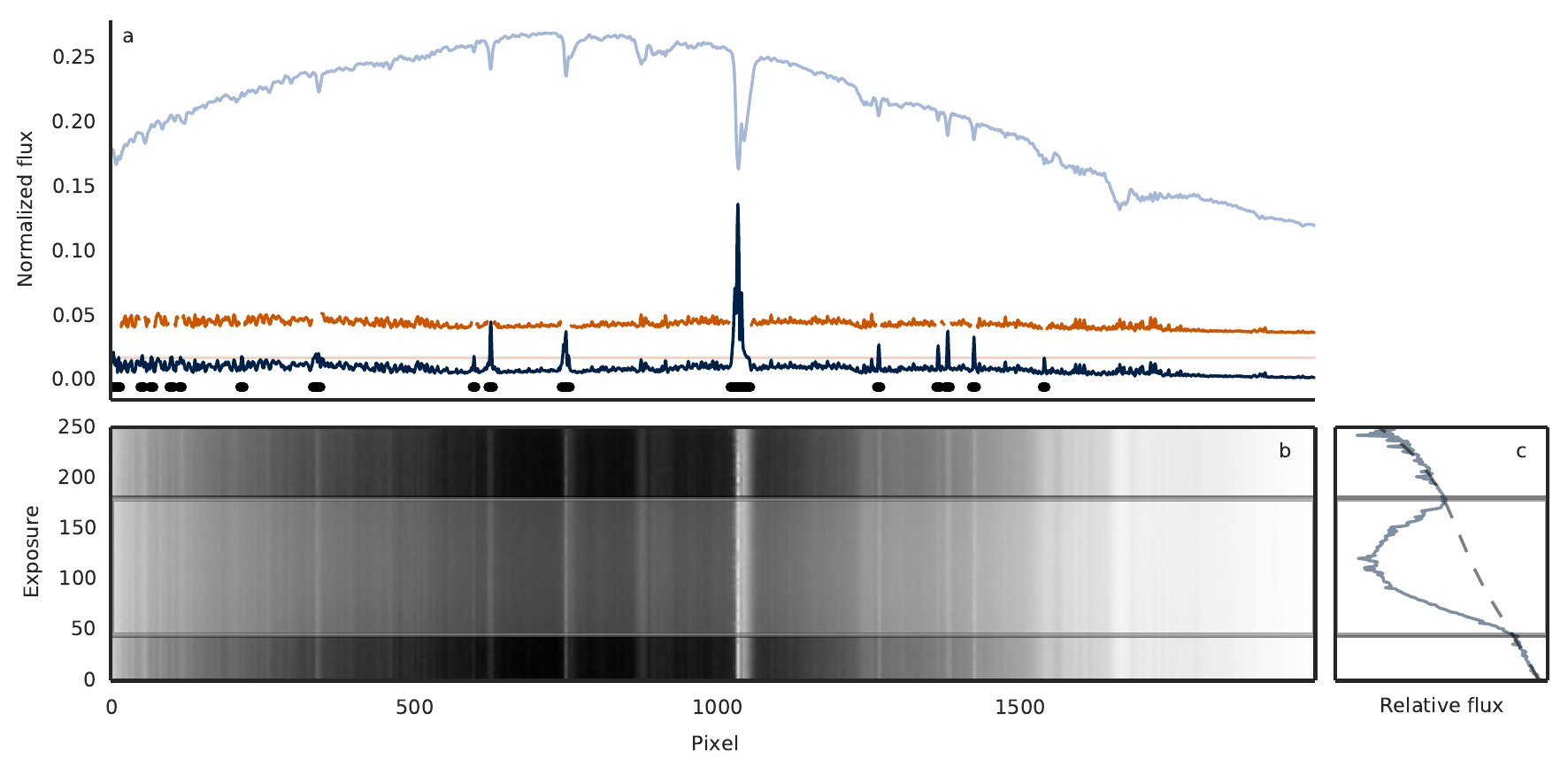}
 \caption{\changed{Spectrum masking: a) TrES-3 spectrum (light-blue line), out-of-transit (OOT) standard deviation
($\sigma_\mathrm{O}$) of a detrended light curve created using a single spectral pixel (dark-blue line), maximum
allowed $\sigma_\mathrm{O}$ (light-orange horizontal line), OOT standard deviation with a mask (orange line), masked
locations (thick black lines in the bottom); b) the whole spectroscopic time series, time in the y-axis and
wavelength (without wavelength calibration) in the x-axis, inside-transit duration marked between the two horizontal
lines; c) broadband light curve with a fourth-order polynomial fitted to the OOT fluxes.}} 
\label{fig:smask}
\end{figure*}

\changed{The masked light curve set was created by first calculating the standard deviations of detrended (using a 
simple fourth-order polynomial) light curves created for each wavelength element (pixel) for TrES-3 and the comparison 
star (Fig.~\ref{fig:smask}). Next, we masked the wavelength elements where the noise level is higher than 
a given maximum noise level, and then proceeded as with the generation of the basic light curve set.}

\begin{figure*}
 \centering
 \includegraphics[width=\textwidth]{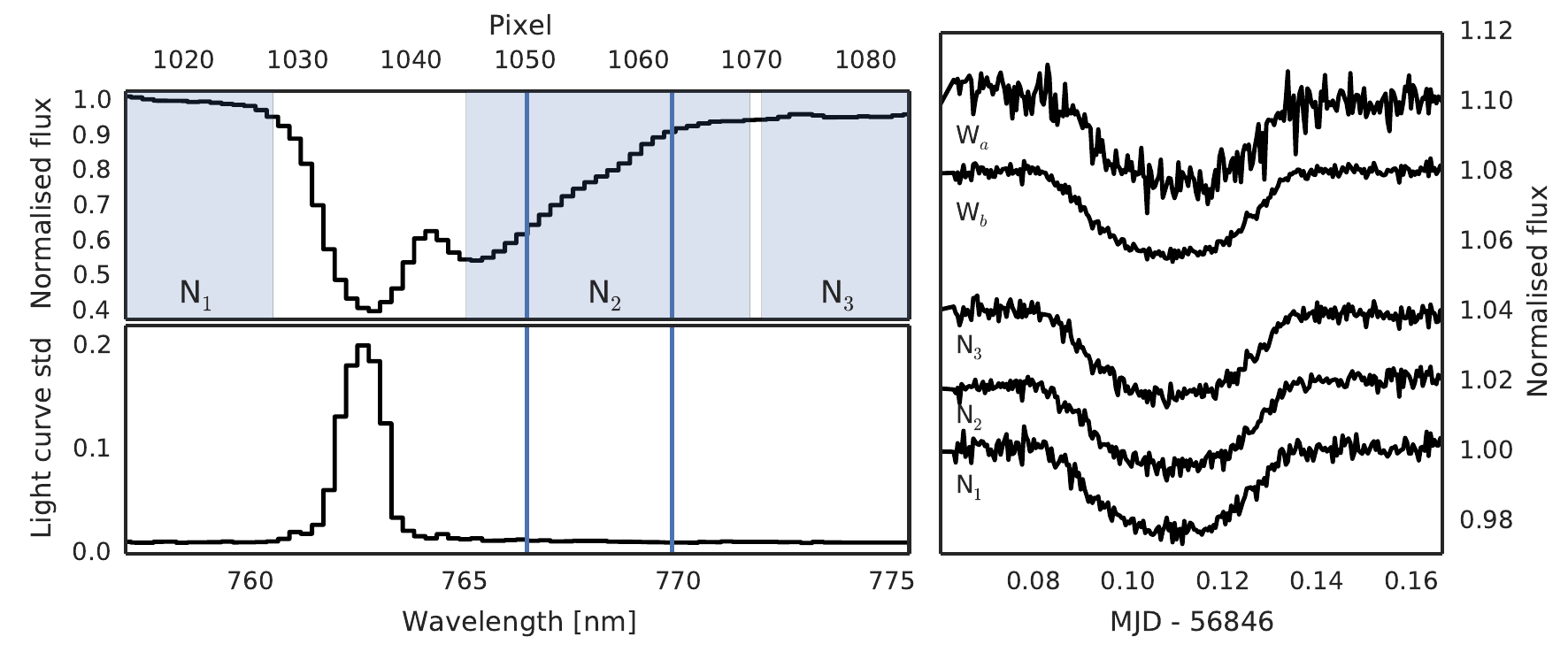}
 \caption{Upper-left panel: a close-up to TrES-3 spectrum with the K~I resonance double lines marked as two vertical 
blue lines, and three passbands used to generate the narrow-band light curves in the panel on the right marked as 
light-blue shaded areas. Lower-left panel: a scatter map based on light curves generated using one pixel in the 
wavelength axis. Right: Light curves generated by integrating over the whole spectral range shown in the left panels 
(W$_a$), integrating over the three marked passbands (W$_b$, that is, excluding the core of the telluric O$_2$ 
absorption band), and integrated over the individual passbands (N$_1$--N$_3$).}
 \label{fig:smask_zoom}
\end{figure*}

\changed{Spectrum masking was found to decrease the rms scatter in the narrow-band light curves covering telluric 
absorption bands, as illustrated in Fig.~\ref{fig:smask_zoom} for the strong O$_2$ absorption band, and the masked light 
curve set was adopted as the main 
analysis dataset.}

\subsubsection{Final transit light curves}

The final light curve sets were created by dividing the \mbox{TrES-3} light curve sets (unmasked and masked) by the 
comparison star light curves. The rms scatter for the broadband curve is~$\sim$500~ppm, with a white noise estimate 
(using Gaussian processes with an exponential kernel to model the time-correlated noise) of~$\sim$350~ppm.

\subsection{Background Contamination}
\label{sec:data:contamination}

\begin{figure}
 \centering
 \includegraphics[width=\columnwidth]{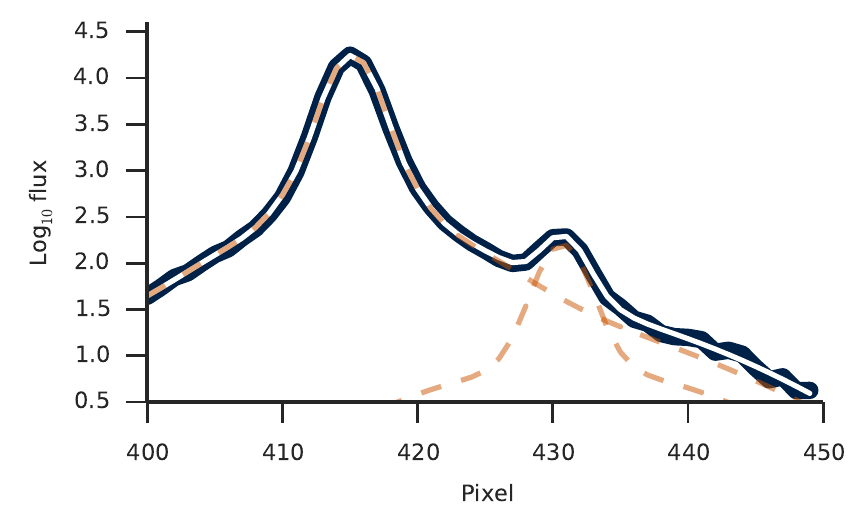}
 \caption{Estimation of the contamination from the faint background star within the TrES-3 aperture. The thick black 
line 
shows the observed flux, the white line the fitted model, and the dashed orange lines the two model components.}
 \label{fig:contamination_model}
\end{figure}

The aperture used to calculate the flux of TrES-3 includes a faint background star (2MASS 17520839+3732378, V=18.5).
The star was estimated to contribute $\sim$1\% of the total flux using PSF fitting. The fitting was done 
using four and three Gaussian components for the TrES-3 and the contaminant PSFs, respectively, as shown in 
Fig.~\ref{fig:contamination_model}. The contaminant is slightly redder than TrES-3b ($J-K = 0.8$, while for TrES-3
 $J-K=0.4$), and we give a (very conservative) constraints on its effective temperature to lie between 3500 and 
5000~K. This temperature is used as a uniform prior later in the analysis.

\section{Theory and numerical methods} 
\label{sec:theory}
\subsection{Overview}
\label{sec:theory:overview}

The analysis consists of a basic broadband parameter estimation run carried out as a consistency test, and a 
transmission spectroscopy run for the 16 narrow passbands shown in Fig.~\ref{fig:spectra}. The approach for the 
parameter estimation is Bayesian: we use Markov chain Monte Carlo (MCMC) to obtain a representative sample from a model 
parameter posterior distribution, where the model jointly describes the observed light curves and a stellar limb 
darkening profile created using the PHOENIX stellar atmosphere library by \citet{Husser2013}. 

The model is parametrised with a parameter vector \pvec. A set of model parameters are passband-independent by nature, 
such as the transit centre and the impact parameter, and each passband adds a set of passband-dependent parameters, all
listed in Table~\ref{tbl:parameters} with their priors. Thus, the number of parameters varies from 10 (broadband light 
curve assuming white noise), to $\sim$100 (16 narrow passbands, red noise). 

\begin{table}[t]
 \caption{Parametrisation and priors. The uninformative priors are uniform and wide enough not to affect
          the posteriors. \NP{\mu,\sigma} stands for a normal prior with mean $\mu$ and standard deviation $\sigma$.
          The colour dependent parameters are per passband.}
 \centering
 \begin{tabular*}{\columnwidth}{@{\extracolsep{\fill}}llll}
  \toprule\toprule
  \multicolumn{2}{l}{Notation} & Name & Prior \\
  \midrule
  \multicolumn{4}{l}{\textit{Informative priors}} \\
  \pper & $P$    & orbital period & \NP{1.306, 10^{-7}} \\
  \\
  \multicolumn{4}{l}{\textit{Uninformative priors, colour independent}} \\
  \ptc  & $T_c$       & transit centre & \\
  \prho & $\rho_\star$ & stellar density & \\
  \pip  & $b$ & impact parameter & \\
  $\theta_\mathrm{Go}$ & $e$ & GP output scale & \\
  $\theta_\mathrm{Gi}$ & $l$ & GP input scale & \\
  \\
  \multicolumn{4}{l}{\textit{Uninformative priors, colour dependent}} \\
  \prr  & $k$ & radius ratio & \\
  \ldu  & $u$ & limb darkening u & \\
  \ldv  & $v$ & limb darkening v & \\
  \pbl  & $B$ & baseline level & \\
  \pext & $x$ & extinction coefficient & \\
  $\theta_\epsilon$ & $\epsilon$ & average white noise & \\
  \bottomrule
 \end{tabular*}
 \label{tbl:parameters}
\end{table}

\changed{The parameter estimation from the narrow-band light curves was carried out separately for three noise-models:
WN) white and normally distributed noise; DW) red noise with a wavelength-independent systematic component; and GP) red
noise described by a Gaussian process with time as the only input parameter.\footnote{We tested for correlations between
the photometry and auxiliary information obtained simultaneously with the observations (airmass, temperature, rotator
angle, etc.), but did not find significant correlations. Thus, we decided not to include any of these as the GP input
parameters \citep[as done in ][for example]{Gibson2011a} due to the added complexity from the increased number of GP
hyperparameters} The broadband modelling used only WN and GP approaches. The likelihoods equations for each of these
cases are detailed in Sect.~\ref{sec:analysis:posteriors}. }

Our parameter estimates correspond to the posterior median, and the uncertainties correspond to the 68\% 
central posterior intervals, unless otherwise specified. We adopt the red noise results as our final results, due to 
their robustness over the white-noise assumption, but also describe any differences between the two.
 
The analysis relies on Python- and Fortran-based code utilising SciPy, NumPy \citep{VanderWalt2011}, IPython
\citep{Perez2007}, Pandas \citep{Mckinney2010}, matplotlib \citep{Hunter2007},
seaborn,$\!$\footnote{\url{http://stanford.edu/~mwaskom/software/seaborn}} PyFITS,$\!$\footnote{PyFITS is a product of
the Space Telescope Science Institute, which is operated by AURA for NASA} and F2PY \citep{Peterson2009}. The transits
were modelled with PyTransit\footnote{Freely available from \url{https://github.com/hpparvi/PyTransit}}
\citep{Parviainen2015}, \changed{the limb darkening computations were carried out with PyLDTk,}$\!$\footnote{Available
from \url{https://github.com/hpparvi/ldtk}} global optimisation was carried out with PyDE,$\!$\footnote{Available from
\url{https://github.com/hpparvi/PyDE}} the MCMC sampling was carried out with emcee
\citep{Foreman-Mackey2012,Goodman2010}, and the Gaussian Processes were computed using George\footnote{Available from
\url{https://dan.iel.fm/george}} \citep{Ambikasaran2014}.

\subsection{Limb darkening}
\label{sec:analysis:limb_darkening}

TrES-3b presents a nearly grazing transit, and its inclined orbit leads to a degeneracy between the planet-star radius 
ratio, impact parameter and stellar limb darkening. When observing TrES-3b, \citet{O'Donovan2007} used multicolour 
photometry to break the degeneracy between the radius  ratio and the impact parameter, allowing them to constrain the 
two, but they fixed the limb darkening coefficients to  values tabulated in \citet{Claret2004}. These tabulations have 
been shown to disagree with observed limb darkening profiles \citep{Claret2008,Claret2009}, and fixing the limb 
darkening coefficients to the tabulated values can lead to biased parameter estimates with underestimated 
uncertainties. Nowadays, more robust approaches to account for limb darkening are widely used. The tabulations can be 
used to construct informative priors on the limb darkening coefficients, where the prior widths depend both on how much 
we trust the stellar models behind the tabulations and how well the host star has been characterised. The limb darkening
coefficients can also be completely unconstrained in the parameter estimation, with uninformative priors, which leads to
the most conservative parameter 
estimates.

We use an approach where, instead of constraining the limb darkening model coefficients, we constrain the shape of the 
stellar limb darkening profile. This is achieved by fitting a stellar limb darkening profile (with uncertainties 
propagated from the uncertainties in our stellar parameter estimates) created using the specific intensity spectra 
library by \citet{Husser2013} jointly with the light curves. The \citet{Husser2013} library allows us to create limb 
darkening profiles for freely defined passbands, and the profile-based approach reduces the problems 
encountered with the limb darkening coefficient correlation.

\subsection{Transmission spectroscopy}
\label{sec:analysis:spectroscopic}
The number 
of model parameters for the narrow-band analysis is $\sim$100, which means that the size of the parameter vector 
population has to be increased for the affine invariant sampler to work. Also, even then, the autocorrelation length of 
the chains is significantly higher than for the lower-dimensional white-light analysis, and more iterations are 
required 
to obtain a usable set of independent posterior samples. A small run was carried out to test whether fixing the 
GP hyperparameters affects the parameter estimates, and no significant effects were observed.
 
\subsection{Posteriors and likelihoods}
\label{sec:analysis:posteriors}

We model the observed spectrophotometry and the theoretical stellar intensity profiles jointly. Our unnormalised
log posterior density is
\begin{equation}
\ln P(\pvec|D) = \ln P(\pvec) + \llh(\Dlc|\pvec) + \llh(\Dld|\pvec),
\end{equation}
where \pvec is the parameter vector encapsulating all the model parameters, $\ln P(\pvec)$ is the log prior, 
\Dlc is the spectrophotometry data, $\ln P(\Dlc|\pvec)$ is the log likelihood for the photometry, \Dld are the
theoretical limb darkening profiles, and $\ln P(\Dld|\pvec)$ is the log likelihood for the limb darkening
profiles.

Assuming that the uncertainties in the observations are normally distributed, we can write the general log likelihood 
for data $\vec{D}$ given the parameter vector \pvec in a vector form as
\begin{equation}
 \llh(\vec{D}|\pvec) = -\frac{1}{2} \left( n_D \ln 2\pi +\ln|\covmat| +\vec{r}^\mathrm{T} \covmat^{-1} 
\vec{r}\right),
 \label{eq:general_posterior}
\end{equation}
where $n_D$ is the number of datapoints, $\vec{r}$ is the residual vector with elements $r_i = D_i-M(t_i,\pvec)$, 
$M$ is the model, and $\covmat$ is the covariance matrix. 

If the noise can be assumed white (that is, uncorrelated), the covariance matrix is diagonal, and the computation 
of the likelihood is trivial. However, if the noise is correlated, the covariance matrix will have off-diagonal 
elements, and the matrix needs to be inverted for the likelihood evaluation.

\subsubsection{Likelihood for the stellar limb darkening profile}
\label{sec:analysis:limb_darkening_likelihood}
Instead of relying on the tabulated limb darkening coefficients, we model stellar intensity profiles calculated for 
TrES-3 and our passbands using the PHOENIX  stellar atmosphere code jointly with the photometric data. This allows us 
to 
marginalise over the whole limb darkening coefficient space that can explain the theoretical stellar intensity (limb 
darkening) profile, and may yield more robust parameter estimates than by using fitted limb darkening coefficients 
directly.

The stellar limb darkening profile data, \Dld, is constructed from stellar limb darkening profiles calculated using the 
PHOENIX code for 27 stellar parameter sets over 16 passbands (corresponding to the ones in our basic spectroscopic 
analysis) and 75 values of $\mu$ (where $\mu = \cos \gamma = \sqrt{1-z^2}$, $\gamma$ is the foreshortening angle, 
and $z$ is the projected distance from the centre of the stellar disk divided by the stellar radius). Let $I_{i,j}$ be 
the mean stellar intensity (averaged over the different stellar parameter sets) for passband $i$ and $\mu_j$, and 
$\sigma_{I,i,j}$ the corresponding standard deviation (uncertainty) of the stellar intensity, and $\epsilon$ a 
multiplicative factor $\geq 1$ to account for the fact that the numerical stellar models used to calculate 
the limb darkening profiles should not be relied on blindly \citep[e.g.,][]{Claret2009}. The uncertainties are
independent, and 
the log likelihood can be written in scalar form as
\begin{equation}
    \llh(\Dld|\pvec) = - \sum_{i=1}^{\npb} \left( \frac{\nip}{2}  \ln 2\pi + \sum_{j=1}^{\nip} 
\ln\epsilon\sigma_{\mathrm{I},i,j} + \frac{\chi^2_i}{2\epsilon^2} \right),
\end{equation}
where
\begin{equation}
 \chi^2_i = \sum_{j=1}^{\nip} 
\frac{\left(I_{i,j}-M_\mathrm{I}(\mu_j,\pvec)\right)^2}{\sigma_{\mathrm{I},i,j}^2},
\end{equation}
\npb is the number of passbands, \nip the number of $\mu$-datapoints per passband (in this case a 
constant), and $M_\mathrm{I}$ is the limb darkening model. 

We chose to use the quadratic limb darkening model \citep{Mandel2002,Gimenez2006}
\begin{equation}
  M_\mathrm{I}(\mu, \pvec) = 1 -\ldu(1-\mu) - \ldv(1-\mu)^2 
\end{equation}
 after running test simulations with a quadratic and a general four-parameter limb darkening model \citep{Gimenez2006}.
The use of four-parameter model did not affect the parameter estimates (within the estimate uncertainties), but
introduced unjustified complexity to the model with two additional parameters per passband.

\subsubsection{Likelihood for the photometry assuming white noise}
\label{sec:analysis:likelihood_white_noise}

If we assume the noise in the photometry for a single passband to be \iid (independent and identically distributed)
from a zero-centred normal distribution with a standard deviation $\sigma_{\mathrm{lc,i}}$, the likelihood can be
written out explicitly in scalar form as
\begin{equation}
  \llh(\Dlc|\pvec) = -\sum_{i=1}^{\npb} \left( \frac{\nph}{2} \ln 2\pi\sigma_{\mathrm{lc,i}}^2  + \sum_{j=1}^{\nph} 
\frac{\left(F_j-M_\mathrm{P}(\pvec,t_j,X)\right)^2}{2\sigma_{\mathrm{lc,i}}^2} \right)
\end{equation}
where \npb is the number of passbands, \nph the number of photometric datapoints, and $\sigma_{\mathrm{lc,i}}$ the 
average scatter in the $i$th passband.

The photometry is modelled as a product of a baseline and a transit components as
\begin{equation}
M_\mathrm{P}(\pvec, t, X) = \pbl \; e^{-\pext X} \; T(t, \ptc, \pper, \prr, \pa, \pin, \pec, \pom, \ldu, \ldv)
\end{equation}
where \pbl is a constant baseline level, $X$ is the airmass, \pext is the (residual) extinction coefficient, $T$ is the 
transit model, $t$ is the mid-exposure time\footnote{The exposure time is short enough that we do not need to worry 
about the transit shape blurring due to extended integration time}, \ptc is the zero epoch, \pper is the orbital 
period, 
\prr the planet-star radius ratio, \pa the scaled orbital semi-major axis, \pin the inclination, \pec the orbital 
eccentricity, \pom the argument of the periastron, and \ldu and \ldv are the quadratic limb darkening 
coefficients.

The baseline includes a constant baseline level and an atmospheric extinction term to model uncorrected extinction as a 
function of the airmass. The latter term is necessary since TrES-3 and the comparison star have a slightly different
colour. The baseline parameters are passband dependent, and thus yield two free parameters per modelled passband.  The
transit is modelled using \textit{PyTransit}, which is optimised for efficient modelling of spectrophotometric transits.
A part of the transit parameter set is colour independent (transit centre, orbital period, etc.), while the radius ratio
and the limb darkening coefficients are passband dependent.

\subsubsection{Likelihood with red noise modelled with the DW approach}
\label{sec:analysis:likelihood_red_noise_dw}

In reality, the noise in the photometry is rarely white (where with noise we mean the sum of every signal not included
in our model). Instead, we have many time-varying factors affecting our measurements---such as the seeing and the
location of the PSF on the CCD---that introduce systematic signals seen as correlated noise. If these factors are
measured simultaneously with the observations, we can use several approaches to model the signal they add to our
observations, and thus improve the accuracy of our parameter estimates. 

\changed{
The DW approach assumes that the systematic component of the noise is constant (with a possible scaling factor) across
the spectrum. If the assumption holds true, we can model the narrow-band systematic noise with the help of the broadband
light curve, using the ratios of the observed broad- and narrow-band fluxes and the modelled broad- and
narrow-band fluxes. The log likelihood is now
\begin{equation}
 \ln P(D|\pvec) = \ln P(W|\pvec) + \sum_{i=1}^{n_\mathrm{pb}} \ln P(F_\mathrm{i}|\pvec),
\end{equation}
where the first term is the broadband log likelihood (assuming independent and identically distributed noise following
the normal distribution), $W$ is the broadband flux, $F_\mathrm{i}$ is the $i$th narrow-band flux, and the terms inside
the sum are 
\begin{equation}
 \ln P(F_\mathrm{i}|\pvec) = -N \ln \sigma_\mathrm{r} - \frac{\ln 2\pi}{2} - \sum_{j=1}^N 
\frac{\left(\frac{\alpha F_\mathrm{i,j}}{1+\beta\left(W_\mathrm{j}-1\right)} - 
\frac{M_\mathrm{i,j}}{1+\beta\left(M_\mathrm{W,j}-1\right)}\right )^2}{2\sigma_\mathrm{r}^2},
\label{eq:logl_relative_flux}
\end{equation}
where $\sigma_\mathrm{r}$ is the flux ratio 
scatter, $\alpha$ is the constant baseline level for the flux ratio, $M$ is the modelled narrow-band flux,
$M_\mathrm{w}$ is the modelled broadband flux, and $\beta$
is a scaling factor applied to both observed and modelled wide-band fluxes. 
The approach is similar to the often-used method of first fitting the wide passband and subtracting the residuals from 
the narrow-band light curves, but slightly more robust, since we are marginalizing over the baseline and scale 
parameters $\alpha$ and $\beta$, and modelling the relative flux explicitly.}

\subsubsection{Likelihood with red noise modelled as a Gaussian process}
\label{sec:analysis:likelihood_red_noise_gp}

Gaussian processes offer a model-independent stochastic way to include the effects from several sources of systematic
signals \citep{Rasmussen2006,Gibson2011a,Roberts2013}.
The covariance matrix \covmat in Eq.~\eqref{eq:general_posterior} is now
\begin{equation}
 \covmat =  \vec{K}(\vec{x},\vec{x}) + \sigma^2\vec{I},
\end{equation}
where $\vec{K}(\vec{x},\vec{x})$ is defined by a covariance function (kernel).
We chose to use a simple exponential kernel with the mid-exposure time as the only input parameter
\begin{equation}
 k(t_i,t_j) = h^2 \exp\left(-\frac{|t_j-t_i|}{\lambda}\right),
\end{equation}
where $h$ is the GP output scale (defines the standard deviation of the Gaussian Process) and $\lambda$ is the input 
scale. The likelihood is now given by Eq.~\eqref{eq:general_posterior}, but the full covariance matrix needs to be
inverted. The covariance matrix is symmetric and positive semi-definite, which ensures that the inversion is always
possible, but the inversion is still numerically costly. 

We marginalise over the GP hyperparameters in the white-light curve analysis. For the spectroscopic analysis, we first
optimise the GP hyperparameters to the white-noise analysis residuals. We assume that the GP hyperparameters are
passband-independent, but the white noise component varies form passband to passband.

\section{Broadband analysis}
\label{sec:broadband_analysis}
\subsection{Overview}
\label{sec:broadband:overview}

\begin{figure}
 \centering
 \includegraphics[width=\columnwidth]{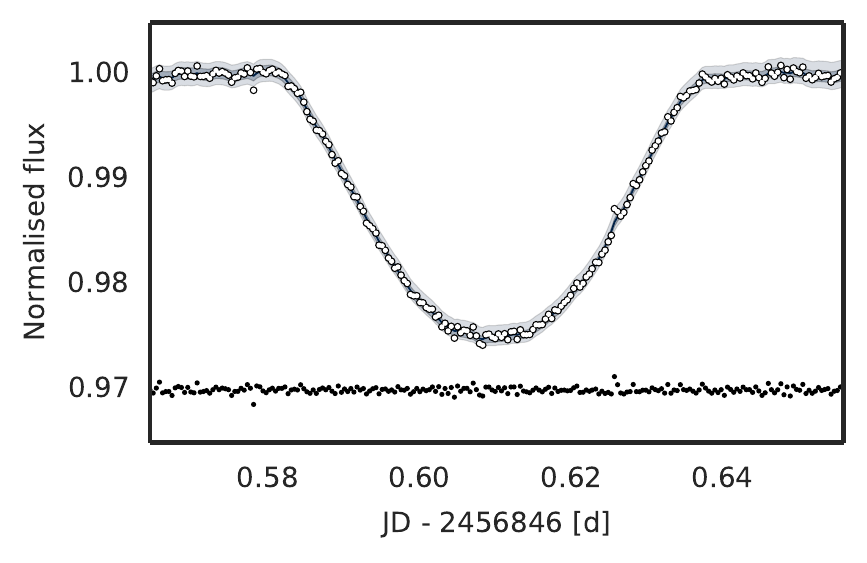}
 \caption{Observed white light curve (white points), the 68\% and 95\% central intervals of the conditional model
distribution assuming red noise (light and dark grey), and the residuals (black points).}
 \label{fig:white_lc_and_model}
\end{figure}

\begin{figure*}
 \centering
 \includegraphics[width=\textwidth]{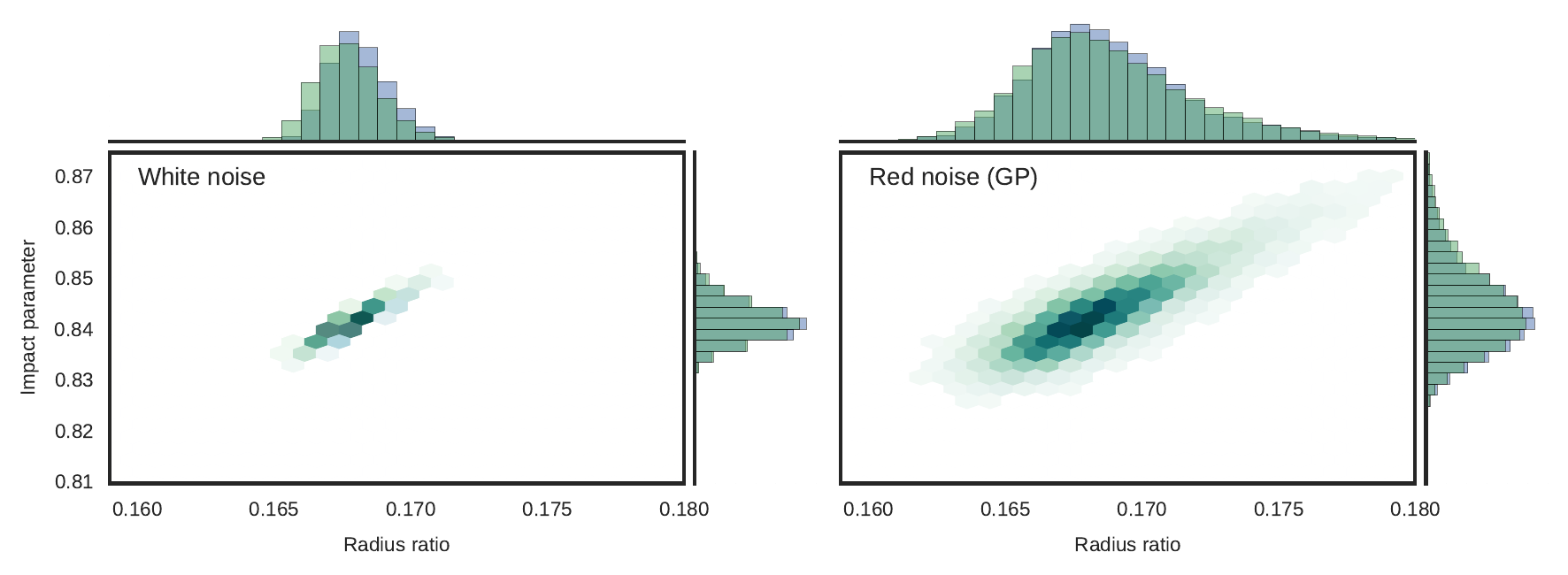}
 \caption{Joint posterior distributions for the radius ratio and impact parameter for the four broadband analyses using 
unmasked (blue) and masked (green) light curves assuming either white (left) or red (right) noise.}
 \label{fig:white_k_and_b}
\end{figure*}

\changed{We carry out a broadband (white light) analysis as a consistency test, motivated by the possible issues caused 
by the TrES-3b's large impact parameter, and by the need to test whether the spectrum masking has an effect on 
broadband parameter estimates.} TrES-3b has been observed extensively 
\citep{O'Donovan2007,Winn2008,Sozzetti2009,Gibson2009,Ballard2009,Colon2010,Lee2011b,Turner2012,Vanko2013}, but many of
the analyses have imposed strict priors on the limb darkening, or have considered white noise only. However, the radius
ratio, orbital impact parameter and stellar limb darkening are all degenerate, and the transit shape (especially when
allowing for red noise) can be explained by a large poorly constrained subvolume of the parameter space.

\changed{The broadband modelling is carried out for the light curves with and without spectrum masking, assuming either 
white or red noise. The red noise is modelled using a Gaussian process with time as the only input parameter (see 
Sect.~\ref{sec:analysis:likelihood_red_noise_gp}). An exponential kernel producing once-differentiable functions 
\citep{Roberts2013,Rasmussen2006} was chosen over the often-used squared exponential (SE) kernel (producing smooth 
infinitely-differentiable functions) and the slightly more complex Mat\'ern kernel since it was observed\footnote{The 
maximum likelihood for the exponential kernel, when fitted to the white-noise-run residuals, was higher than for the SE 
kernel.} to reproduce the noise characteristics better than the SE kernel, but without the additional hyperparameter of
a Mat\'ern kernel. We marginalise over the two GP hyperparameters, the length scale and output scale.}

\changed{The parameter estimation for all cases starts with a parameter vector population that fills uniformly the 
prior space. An initial differential evolution (DE) optimisation is used to clump the population close to the global 
posterior maximum, after which MCMC sampling is carried out using \textit{emcee}. The sampler is run for 10$\,$000 
iterations, which yields 9000 independent posterior samples (using a population size of 100, thinning factor of 100, 
and burn-in period of 1000 iterations, where the thinning factor and burn-in period have been chosen by studying the 
chain population.)}

\subsection{Results}

\changed{The observed light curve, conditional model distribution (for the red noise model) and the residuals are shown 
in Fig.~\ref{fig:white_lc_and_model} and the joint posterior distributions for the radius ratio and impact parameter in 
Fig.~\ref{fig:white_k_and_b}. The broadband analysis results agree with all the previous TrES-3b analyses, 
but we will not report the numerical estimates here. Simultaneous multicolour transit modelling described in the next 
section alleviates the degeneracies between the parameters, and we will adopt the narrow-band analysis results as our 
final parameter estimates. No significant discrepancies were identified between the masked and unmasked results.}

\section{Transmission spectroscopy}
\label{sec:transmission_spectroscopy}
\subsection{Overview}
\label{sec:tran:overview}

\changed{The transmission spectroscopy is carried out for masked and unmasked light curves and three approaches to 
modelling noise, and follows closely the broadband analysis. The main difference is the need to use a larger parameter 
vector population (due to high model dimensionality), and special care is needed to ensure that the sampler has 
converged to sample the true posterior distribution. An initial population of 300 (white noise) or 400 (red noise) 
parameter vectors is clumped around the global posterior maximum using the DE algorithm, and the population is then used 
to initialise the MCMC sampler. The MCMC sampling was carried out repeatedly over 15$\,$000 iterations, each run 
starting from the last iteration of the previous run, until the per-run parameter medians were stable (did not show 
significant trends compared to the parameter vector population scatter) over the run. The mean autocorrelation lengths 
were estimated from the MCMC chains, and a thinning factor of 100 was used to ensure that the samples are not 
significantly correlated. }

\changed{The red noise model uses GPs similarly to the broadband analysis. Now, however, we do not marginalise over the 
GP hyperparameters, but fix them to values optimised before the MCMC run (against the residuals from the white-noise 
run). We tested whether this affects the parameter estimates with a short MCMC run with free GP hyperparameters, but 
did not find any significant differences.}

\subsection{Results}
\label{sec:tran:results}

\begin{figure*}[t]
 \centering
 \includegraphics[width=\textwidth]{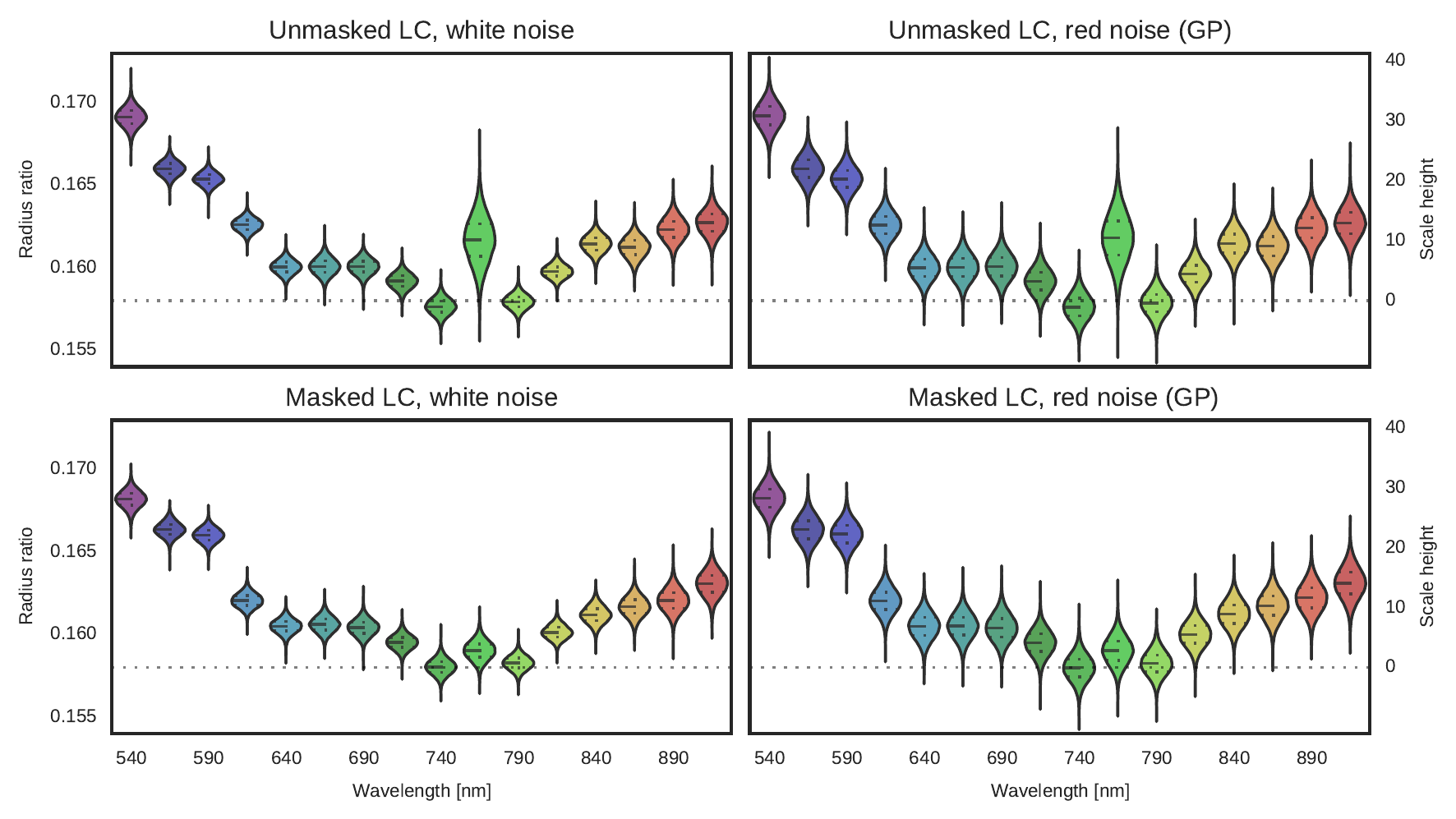}
 \caption{TrES-3b transmission spectrum assuming white noise (left) or red time-correlated noise (right) for 25~nm 
spectral bins covering 530~nm to 930~nm. The potassium doublet lines are at 766.5~nm and 769.9~nm}
 \label{fig:transmission_spectrum}
\end{figure*}

\begin{figure*}[t]
 \centering
 \includegraphics[width=\textwidth]{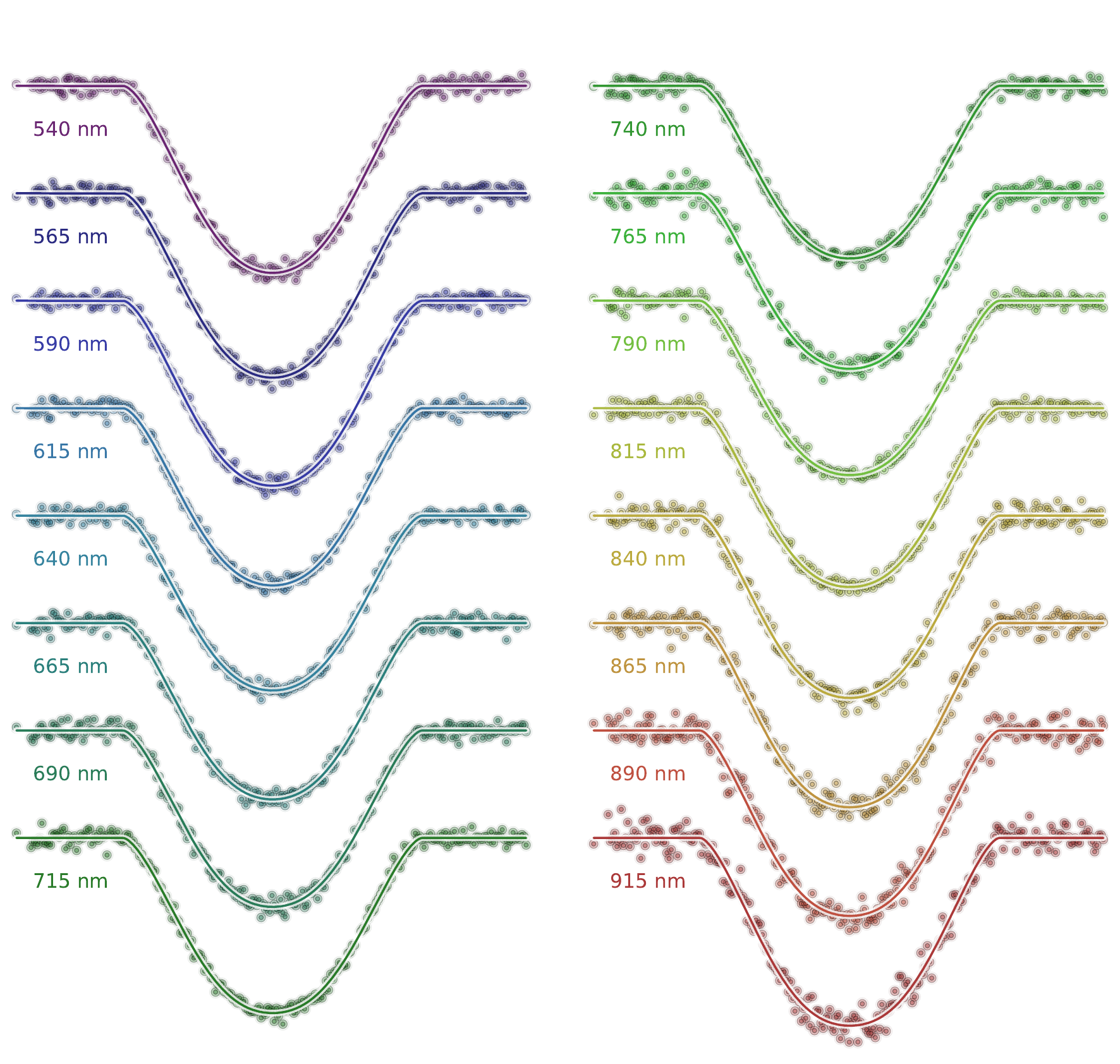}
 \caption{Observed spectrophotometry with spectrum masking (points) and the fitted model for the 25~nm spectral bins. 
We have subtracted the GP prediction mean from the observed datapoints and divided by the baseline model for 
visualisation purposes. A version without the subtraction of the GP prediction mean can be found from the supporting 
IPython notebook.}
 \label{fig:lc_spec}
\end{figure*}

\changed{
We report the wavelength-independent parameters in Table~\ref{tbl:final_estimates}, and show the derived transmission 
spectra for the 16 narrow-band light curves spanning 530~nm to 930~nm in 25~nm bins in 
Fig.~\ref{fig:transmission_spectrum}, and the narrow-band light curves (with spectrum masking) and the model in 
Fig.~\ref{fig:lc_spec}. Shown radius ratio estimates are relative to the average radius ratio. The absolute radius 
ratio estimates include an uncertainty in the average radius ratio, which will be a major factor in the uncertainty of 
the per-passband radius ratio estimates. In transmission spectroscopy we are interested in the relative changes 
between the passbands, and the uncertainty in the average radius ratio is not of interest. The estimates shown here have 
been corrected for this absolute shift by dividing the radius ratios of each MCMC sample  with their average, and 
then multiplying the relative radius ratios with the total posterior sample mean (0.162). We omit the results from the 
divide-by-white analysis. The parameter estimates from it are close to the white noise results, only with slightly 
reduced uncertainties.}

\changed{The transmission spectrum from the unmasked light curves features a steep increase in radius towards the blue 
end of the spectrum, and a single peak near 775~nm. The increase towards the blue corresponds 
to $\sim$30 atmospheric scale heights, which is more than can be realistically expected from Rayleigh scattering 
(we assume planetary equilibrium temperature of 1620~K and \logg{} of 3.45.) The 767~nm bin that includes the K~I 
resonance doublet also stands out. However, the bin also covers a strong telluric O$_2$ absorption band, as was shown 
in Figs.~\ref{fig:smask}~and~\ref{fig:smask_zoom}, and spectrum masking completely removes this signal.}

\begin{figure}[t]
 \centering
 \includegraphics[width=\columnwidth]{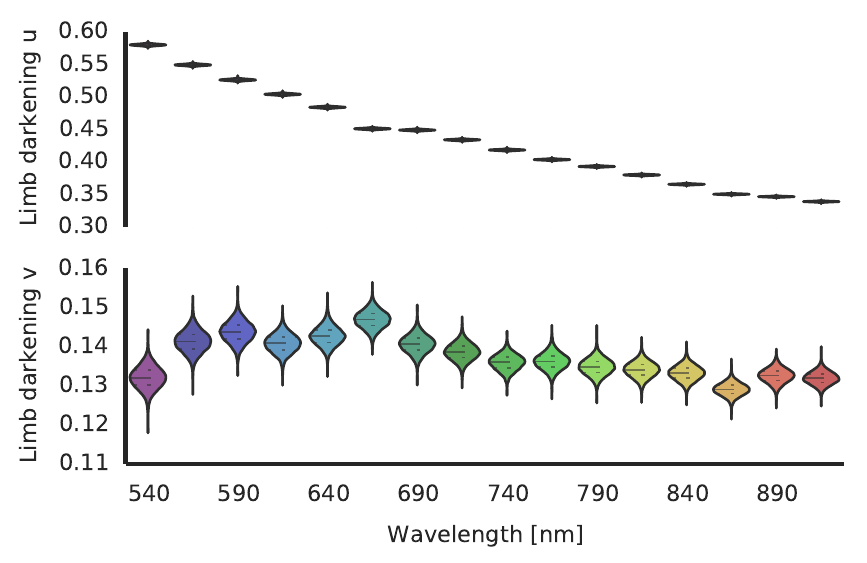}
 \caption{Quadratic limb darkening coefficients $u$ and $v$  for the 25~nm spectral bins covering 530~nm to 930~nm. 
Note that the y-axis scale is different for $u$ and $v$.}
 \label{fig:spec_limb_darkening}
\end{figure}

The estimated quadratic limb darkening coefficients are shown in Fig.~\ref{fig:spec_limb_darkening}. The posterior 
estimates are dominated by the likelihood from the stellar limb darkening profiles, since the nearly grazing orbit 
makes limb darkening poorly constrained by the photometry.

\begin{figure}[t]
 \centering
 \includegraphics[width=\columnwidth]{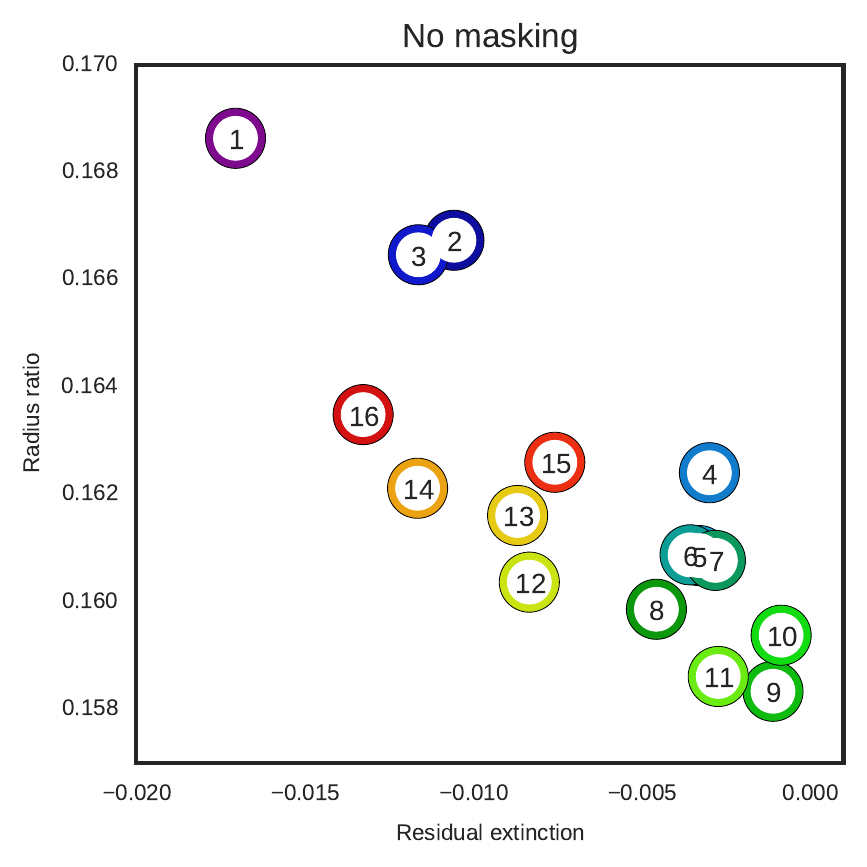}
 \caption{Narrow-band radius ratios as a function of residual extinction. The numbers indicate the passband, starting 
from the shortest wavelength.}
 \label{fig:k_vs_x}
\end{figure}

\changed{Finally, Fig.~\ref{fig:k_vs_x} shows the narrow-band radius ratios as a function of residual extinction 
coefficient estimates. The residual extinction coefficients model the atmospheric extinction that is not corrected by 
dividing the TrES-3b light curves with the comparison star light curves due to different stellar types (that is, the 
spectra of the two stars are different.) The two parameters are correlated (with a correlation coefficient of -0.83), 
but it is difficult to assess whether the correlation implies causation. Rayleigh scattering in the Earth's atmosphere 
leads to stronger extinction in blue, and a similar scattering mechanism in the Planet's atmosphere could in theory be 
behind the observed increase in the radius ratio.}

\begin{table}[t]    
  \caption{Final parameter estimates from the narrow-band analysis.}
  \centering
  \begin{tabular*}{\columnwidth}{@{\extracolsep{\fill}}llr@{$\quad\pm\;$}r}
  \toprule\toprule
  Parameter & Unit & \multicolumn{2}{c}{Value} \\
  \midrule              
  Transit centre    & [MJD]    & 56846.10945 & 4$\times$10$^{-5}$ \\
  Stellar density   & [\gcm]   &       2.370 & 3.4$\times$10$^{-2}$\\
  Impact parameter  &          &       0.844 & 4$\times$10$^{-3}$ \\
  Mean radius ratio &          &       0.162 & 1$\times$10$^{-3}$ \\
  \bottomrule
  \end{tabular*}
  \label{tbl:final_estimates}  
\end{table}

\subsection{Rayleigh-like signal}
\label{sec:tran:bluewards_signal}
    
The TrES-3b's radius ratio increases rapidly from 645~nm towards bluer passbands. \changed{This signal is similar to 
what could be expected from Rayleigh scattering, but its amplitude, about 30 atmospheric scale heights assuming 
planetary equilibrium temperature of 1620~K, is significantly larger than expected from theoretical considerations. The 
amplitude in scale heights would decrease were the temperature of the observed atmospheric layer to be significantly 
higher than the planetary equilibrium temperature, and we address this in the end of 
Sect.~\ref{sec:tran:bluewards_signal}.
}

The planet's atmosphere is not the only factor affecting the transmission spectrum, but other sources can have a
wavelength-dependent effect on the radius ratio estimates. Especially, systematic errors in the limb darkening models, 
unocculted star spots, occulted plages, and contamination from an unresolved source can all lead to colour-dependent 
signals in the radius ratio estimate. Here we first study the effects from each of these one by one, and then combine
them into a toy-model to test whether Rayleigh scattering combined with unocculted spots and unaccounted-for
contamination could explain the spectrum.
    
\subsubsection{Rayleigh scattering with a constant cloud-deck}
\label{sec:discussion:rayleigh}

Considering only Rayleigh scattering, the slope for the planetary radius is
\begin{equation}
    \frac{\ud R_{\rm p}}{\ud \ln \lambda} = \frac{\alpha T k_\mathrm{B}}{\mu g} = \alpha H,
\end{equation}
where $\alpha = -4$, $H$ is the atmospheric scale-heigh, $T$ is the effective temperature, $k_\mathrm{B}$ is the
Boltzmann
constant, $\mu$ is the mean molecular weight of the scatterer, and $g$ is the planet's surface gravity. The equation
for the slope in 
planet-star radius ratio is now
\begin{equation}
    \frac{\ud k}{\ud \ln \lambda} = \frac{\alpha H}{\rstar},
\end{equation}
from where we get a simple model for the radius ratio
\begin{equation}
    k = \frac{\alpha H}{\rstar} \ln \lambda + C,
\end{equation}
where  $C$ is a constant. If we choose $C$ so that the Rayleigh-scattering $k$ 
intercepts the constant cloud deck at $k_0$ for the wavelength $\lambda_0$, we get
\begin{equation}
 C = k_0 -\frac{4H}{\rstar}\ln\lambda_0,
\end{equation}
and the toy-model becomes
\begin{equation}
    \kreal = \max\left( k_0 - \frac{4H}{\rstar} \left ( \ln \lambda  + \ln\lambda_0 \right ), k_0 \right).
\end{equation}
Adopting $T_\mathrm{eq} = 1623 \pm 26$~K and $\log g_\mathrm{p} = 3.45 \pm 0.02$ from \citet{Torres2008}, and $\mu$ as
2.3 time the proton mass, we obtain a normal prior $H \sim N(\text{mean}=205~\mathrm{km}, \text{std}=16~\mathrm{km})$.

\subsubsection{Effects from unocculted spots}
\label{sec:discussion:unocculted_spots}

The presence of unocculted star spots can produce an increase in the observed transit depth (and radius ratio) towards 
blue wavelengths (occulted spots would have their effect also, but they could be distinguished from the photometry). 
The observed radius ratio, \kobs, in the presence of unocculted spots can be expressed as
\begin{equation}
 \kobs = \kreal \sqrt{\frac{1}{1-fA_\lambda}}, \qquad A_\lambda = 1 - \frac{P(\lambda, 
\teff-\Delta \tspot)}{P(\lambda,\teff)},
\end{equation}
where \kreal the true geometrical radius ratio, $f$ the spot filling factor, and $A_\lambda$ the wavelength-dependent 
contrast ratio, $P$ is the Planck's law, $\Delta \tspot$ is the difference between the spot temperature and 
effective stellar temperature and \teff is the effective stellar temperature \citep[see][for an in-depth 
treatise]{Ballerini2012}. 

\begin{figure}[t]
 \centering
 \includegraphics[width=\columnwidth]{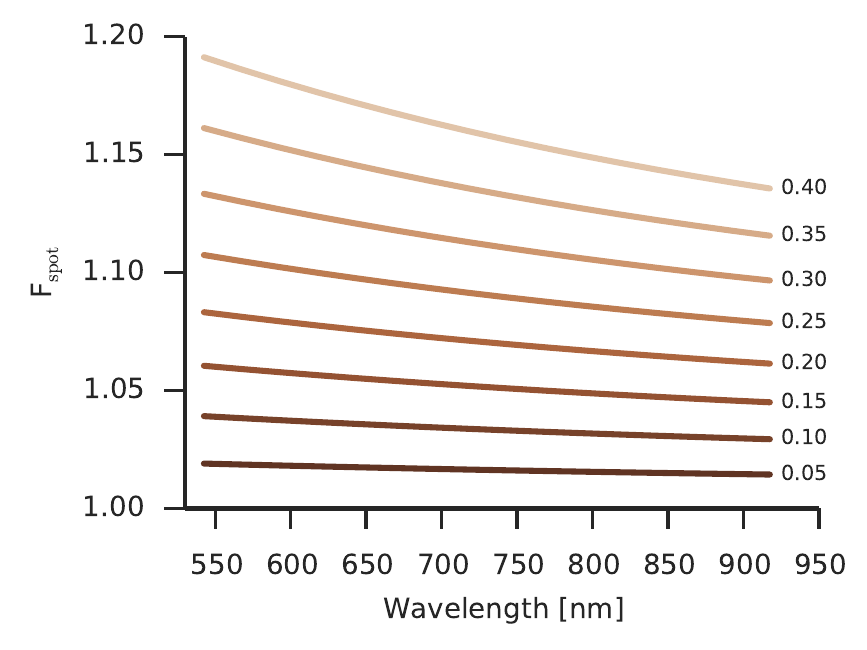}
 \caption{The effect from unocculted spots on the observed radius ratio as a function of wavelength for $\kreal=0.165$, 
$\Delta \tspot = 1200$~K, and filling ratios from 0.05 to 0.40. Unocculted spots always increase the observed radius 
ratio.}
 \label{fig:unocculted_spots}
\end{figure}

As shown in Fig.~\ref{fig:unocculted_spots}, the colour-dependent effect from unocculted spots on \kreal for realistic 
$\Delta \teff$ is nearly linear in visible light, and cannot reproduce the observed signal alone. However, 
variations in the star spot coverage may explain part of the discrepancy between our average radius ratio estimate and 
the previous, smaller, estimates.

\subsubsection{Effects from contamination}
\label{sec:discussion:contamination}

\begin{figure}[t]
 \centering
 \includegraphics[width=\columnwidth]{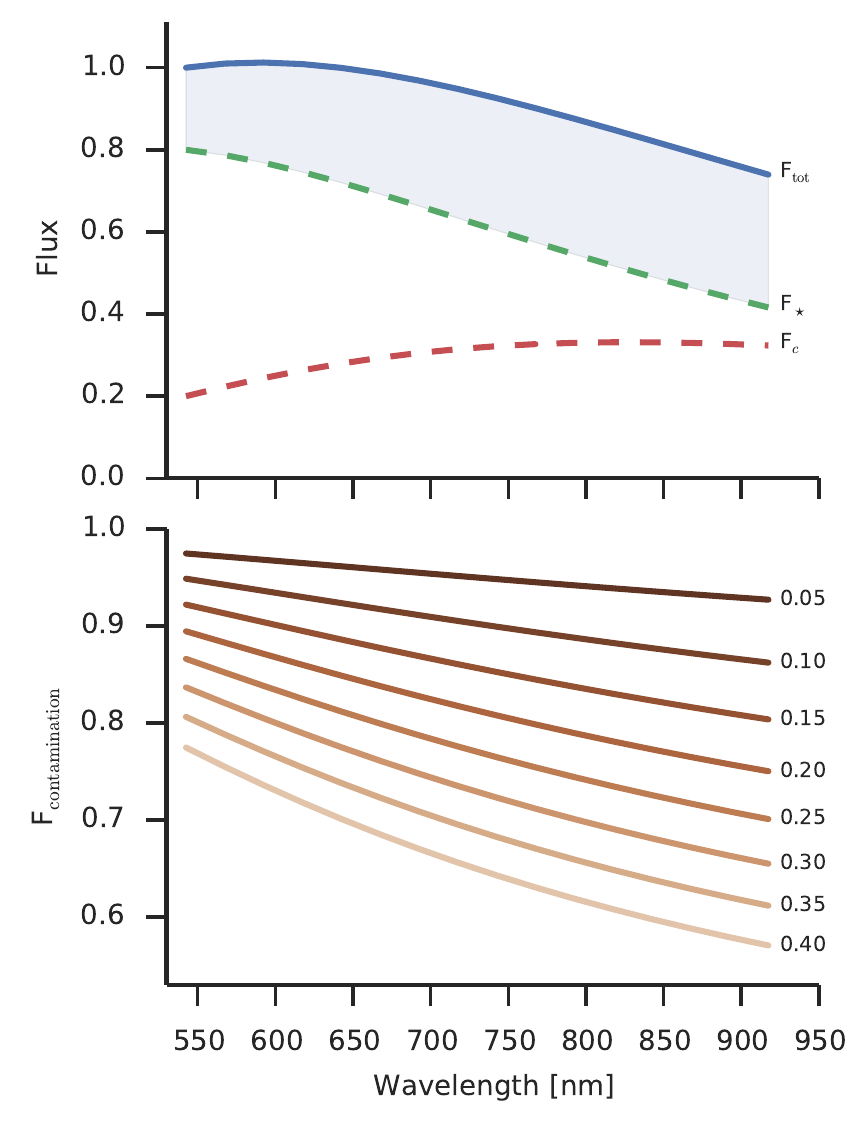}
 \caption{The effect from third-light contamination on the observed radius ratio. The upper panel depicts the
contamination (with an exaggerated 
situation) as light-blue filled region, with a 3500~K contaminant contributing 20\% of the total flux at 545~nm. The 
solid blue line shows the total flux, the upper dashed line the target flux, and the lower dashed line the 
contaminating flux. The lower panel shows the wavelength-dependent effect from contamination for a 3500~K contaminant 
and contamination factor (at 545~nm) varying from 0.05 to 0.40. Contamination always decreases the observed radius 
ratio, but the exact behaviour depends on the temperature difference between the host star and the contaminant.}
 \label{fig:contamination}
\end{figure}

Contamination from an unresolved nearby star falling inside the photometry aperture will also lead to
wavelength-dependent effects on the observed radius ratio and transit shape \citep{Tingley2004}. The observed radius
ratio, \kobs, is related to the real radius ratio, \kreal, and
wavelength-dependent contamination $c_\lambda$ as
\begin{equation}
 \kobs = \kreal \sqrt{1-c_\lambda},
\end{equation}
where the contamination factor is the fraction of the contaminant flux from the total observed flux. If we approximate
the stellar spectrum with a black body, we get
\begin{equation}
 c_\lambda =  \frac{c_0 P_N(\lambda, \tcont)}{c_0 P_N(\lambda, \teff) + (1-c_0)P_N(\lambda, \tcont)} ,
\end{equation}
where $P_N$ is the Planck's law normalised to a reference wavelength, $\lambda_0$ (that is $P_N(\lambda_0,T)=1$), 
\tcont is the contaminant temperature, \teff is the effective stellar temperature, and $c_0$ is the contamination 
factor for the reference wavelength. The effects form contamination are illustrated in Fig.~\ref{fig:contamination}

\subsubsection{Combined model}
\label{sec:discussion:contamination_and_spots}

Combining the effects from Rayleigh scattering, a constant cloud deck, unocculted spots, and possible third-light 
contamination, we obtain
\begin{equation}
 \kobs =  \max\left( k_0 - \frac{4H}{\rstar} \left ( \ln \lambda  + \ln\lambda_0 \right ), k_0 \right) 
\sqrt{\frac{1-c_\lambda}{1-fA_\lambda}},
\end{equation}
which is now a function of fractional spot coverage, spot temperature difference, effective temperature of the 
contaminating star, and the contamination factor for a given reference wavelength. 

\begin{figure}[t]
 \centering
  \includegraphics[width=\columnwidth]{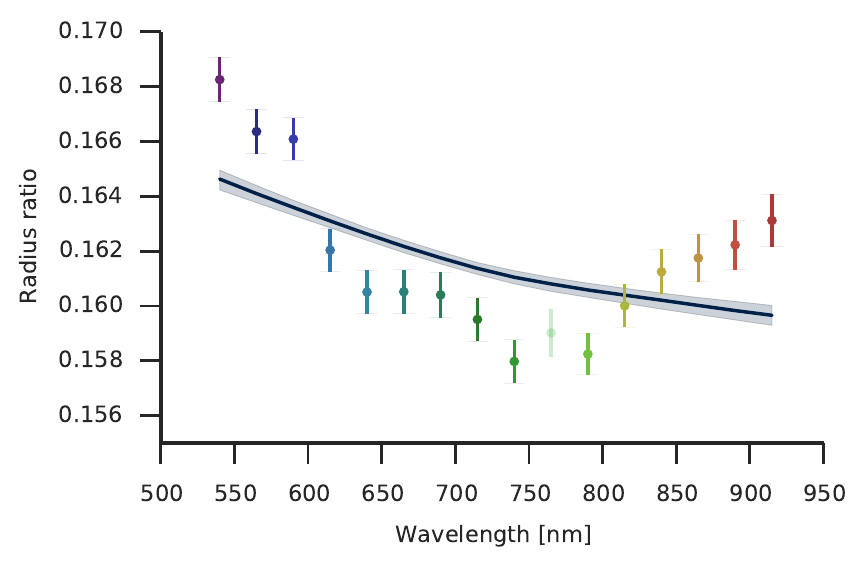}
 \caption{Narrow-band radius ratios and the conditional model distribution for a toy-model that includes Rayleigh 
scattering, flux contamination, and unocculted spots.}
 \label{fig:final_spectrum_model}
\end{figure}

\changed{We carry out an MCMC analysis with a uniform prior on the temperature of the observed atmospheric layer,
ranging from the equilibrium temperature to four times the equilibrium temperature, and show the results in
Fig.~\ref{fig:final_spectrum_model}. Rayleigh scattering is not able to explain the observed slope, even with the
largest possible contributions from unocculted spots and contamination.}

\changed{Finally, we carry out the analysis with an uninformative prior on atmospheric temperature to obtain an estimate 
for the temperature that would be required to explain the slope. We obtain a temperature estimate $T\sim50\,000$~K.}

\subsection{Systematic errors from limb darkening}
\label{sec:spec:limb_darkening}

The stellar limb darkening is one of the main factors affecting the radius ratio estimates, especially with grazing 
orbits. Our approach uses the PHOENIX-generated limb darkening profiles as input information, and if the code were to 
feature systematic deviations from the reality (for example, exaggerated limb darkening towards blue), this would 
directly affect the \prr estimates.

We tested whether the \prr difference of 0.01 between the 542.5~nm and 642.5~nm centred passbands (1st and 5th, 
starting from the bluest) could be explained by systematic errors in limb darkening by first generating a transit light 
curve corresponding to other passband, and then fitting a transit model to this with free limb darkening coefficients 
and \prr set to the other passband value.

The 542.5~nm light curve can also be explained with $\prr = 0.175$ and $u=-0.27$ and $v=0.98$. This leads to negligible 
limb darkening, and is unrealistic at best. The 642.5~nm light curve can be explained with $\prr = 0.185$ and $u=1.1$
and $v=-0.4$. This would mean that all the redwards limb darkening models would significantly underestimate limb
darkening.

For the limb darkening to increase the transit depth when moving towards bluer wavelengths (for a grazing orbit), the 
overall strength of limb darkening would need to decrease with decreasing wavelength (that is, the limb brightness must 
increase). 

We also carry out a parameter estimation run with a constant radius ratio and impact parameter (both constrained with an
informative prior), and limb darkening and the baseline as the only passband-dependent factors. The model fails to
reproduce the variations in radius ratio.

\section{Conclusions}
\label{sec:conclusions}

\changed{We have carried out a transmission spectroscopy analysis for TrES-3b, finding initially a strong Rayleigh-like 
increase in the radius ratio towards the blue end of the spectrum, and a Potassium-like feature near 760~nm. Detailed 
analysis showed that the Potassium-like feature is due to telluric O$_2$ absorption, but the origin of the bluewards 
signal is still unknown (although the correlation with the residual extinction coefficient included into the model 
suggests that it is not a real feature.) We have included possible flux contamination and unocculted spots into a 
simple toy model testing how much of the radius ratio variation could be explained by these effects, but note that they 
fail to explain more than a minor fraction of any wavelength dependent signal.}

\changed{It is clear that more observations are required to test whether the Rayleigh-like signal is a real atmospheric 
feature. Current results do not justify a detailed atmospheric modelling. However, if the strong bluewards signal is 
corroborated by additional observations -- transmission spectroscopy or wide-band photometry -- more serious modelling 
is called for to investigate any possible physical processes behind it. Also, transit observations in the NIR, where 
limb darkening is weaker, will be useful to further constrain TrES-3b's impact parameter.}

\begin{acknowledgements}
We are grateful to Joanna Barstow for her constructive and helpful comments.
HP has received support from the Leverhulme Research Project grant RPG-2012-661. FM acknowledges the support of the
French Agence Nationale de la Recherche (ANR), under the program ANR-12-BS05-0012 Exo-atmos. The work has
been supported by the Spanish MINECO grants ESP2013-48391-C4-2-R and ESP2014-57495-C2-1-R. Based on observations made
with the Gran Telescopio Canarias (GTC), installed in the Spanish Observatorio del Roque de los Muchachos of the
Instituto de Astrofísica de Canarias, in the island of La Palma.
\end{acknowledgements}

\bibliographystyle{aa}
\bibliography{tres_3b}

\end{document}